\documentstyle[12pt]{article}
\begin{document}
\def\in{\parindent=1cm}
\def\no{\parindent=0cm}
\rightline{TUHE9582}
\begin{center}
{\large\bf {Rest Frame Valence Quark Model\\
for Deep Inelastic Scattering}}
\vskip .5cm
J. Franklin and M. Ierano
\vskip .2cm
{\it Department of Physics, Temple University,\\
Philadelphia, Pennsylvania 19122, USA}\\
August 1995
\end{center}
\begin{abstract}

A rest frame valence quark model is applied to the deep
inelastic scattering of charged leptons by protons.  The
parameters of the model are determined by a fit to
unpolarized electron cross sections.  The
model then can be used to calculate the asymmetry in
polarized deep inelastic scattering.  The predicted spin-
dependent structure function, g$_{1p}$, is in good agreement
with recent measurements.  This indicates
that, in the proton rest system, the spin of the proton
is carried by the valence quarks.
\end{abstract}
{\bf 1.  Introduction}
\vskip .5cm

     The deep inelastic scattering of leptons by protons
has been studied since the early 1970's as a tool to
investigate the structure of the proton.  The near
scaling region of about 4 GeV$^{2}<$Q$^{2}<$20 GeV$^{2}$
is of interest because this Q$^{2}$ is large enough to
make the approximation of incoherent scattering by
individual quarks reasonable, and most of the accurate unpolarized
electron data [1-7] and polarized electron\cite{slac,slac2} and muon
results\cite{emc,smc} are for Q$^{2}$ in this range.

    The standard paradigm for analyzing deep inelastic
scattering has been the parton model implemented in the
infinite momentum frame.  By
its nature, the parton model is applicable to the scaling
region as Q$^{2}\rightarrow
\infty$, but is not as
effective in the near scaling region where higher twist
diagrams are required.  The parton model has also led to
the strange interpretation of  recent polarization
experiments that the quark contribution to the
spin of the proton is near zero.

    An alternative method,[12-17] treating the
entire deep inelastic scattering process in the rest
frame of the initial proton, was  introduced some time
ago and shown to be successful in calculating the
approach to scaling of the nucleon
structure functions as Q$^{2}$ increased from 4 GeV$^{2}$
up to q$^2\sim$ 20 GeV$^2$, where  logarithmic QCD corrections
become important.   The
rest frame model  also gave an understanding of the
behavior of F$_{2n}$/F$_{2p}$ at large x as
arising from SU(6) breaking differences in the u and d
quark rest frame wave functions in the     proton, and a
value for R=$\sigma_{L}/\sigma_{T}$  that was
consistent with the early experimental estimates [2].

    In this paper, we apply the rest frame model to the
reanalysis by Whitlow [18] of the early
electron data [1-6], which has resulted
in more accurate combined cross sections with consistent
radiative corrections and a more systematic treatment of
the relative normalizations of the different experiments.
Fitting the rest frame parameters to these electron cross
sections determines the parameters of the model.  We then
make predictions for the asymmetry observed in polarized
lepton-proton deep inelastic scattering, and
for the spin dependent
structure functions, g$_{1p}$ and g$_{2p}$, deduced from
the polarization asymmetry.  We
find good   agreement with the experimental determinations
of g$_{1p}$, which leads us to conclude that, in
the proton rest system, the valence  quarks account for
the proton spin.

     In section 2  of this paper, we describe the rest
frame model for unpolarized deep inelastic scattering.
Section 3   describes the deep inelastic scattering cross
section data, as combined by Whitlow.  Our
fit to these cross sections is presented in Section 4.
In section 5, the rest frame model is extended to the
scattering of polarized leptons by polarized protons, and
the resulting predictions for the asymmetry and spin
dependent structure functions are compared with the polarization data.
Section 6  is a general discussion,
and our conclusion is presented in Section   7.
\vskip  1cm
\no
{\bf 2. The rest frame model}
\vskip .7cm
2.1.  LEPTON-QUARK SCATTERING
\vskip .5cm
\in

     The rest frame model treats deep inelastic
scattering as the quasi-elastic scattering of the lepton
by a point Dirac quark via one-photon
exchange.  The initial quark wave function is that
of a Dirac particle bound in some potential in the
rest frame of the proton.  The final state of the
struck quark is that of a free quark of mass m.
The two spectator quarks will, in general, have
some relative momentum distribution characterizing
their mutual interaction.  In the simple version
of the model used here, this momentum distribution
is taken to be a delta function so that the
spectator quarks appear to be a single diquark of
mass M$_r$.  The cross section for deep inelastic
scattering of a charged lepton from a proton is then
given by the sum of the lepton-quark cross sections,
weighted by the quark charges squared.
The rest frame model is described in more detail, and
contrasted with the usual parton model in I.
In the following we present the equations that
follow from the model.

    Energy conservation in the proton rest system
is given by
\begin{equation}
1+\nu =E'+E_{r},
\label{eq:energy}
\end{equation}
with
\begin{eqnarray}
E' & = & [(\vec{p} +\vec{q})^{2}+m^{2}]^{1/2},\\
E_{r} & = & (\vec{p}^{2} + M_r^{2})^{1/2},
\end{eqnarray}
where q$^{\mu}=(\nu ,\vec{q})$ is the lepton
  four-momentum loss (or the virtual photon
  four-momentum) and $\vec{p}$ is the Fourier
  transform variable for the initial quark wave
  function.  (This can be thought of as the initial
quark momentum.)  We use units in which the proton mass
$M_p$=1.  The recoil energy, E$_r$ of the two spectator
quarks introduces the effective recoil mass M$_r$ as a
parameter of the model.  The photon laboratory energy
$\nu$
can be put into Lorentz invariant form as $\nu=P\cdot q$,
where P$^{\mu}$
 is the initial proton four-momentum, given
in the laboratory by (1,$\vec{0}$).

     The lepton scattering calculation is
straightforward (see the appendix of I).  The
laboratory cross section for the scattering of a charged
lepton of four momentum $k^{\mu}=(\omega,\vec{k})$ into
$k'^{\mu}=(\omega',\vec{k'})$ by a
bound quark  of unit charge is given by
\begin{equation}
{d^{2}\sigma\over{d\Omega d\omega'}}=
\left({\omega'\over{\omega}}\right)
{1\over2\pi^2}\int{d^3p|\phi(p)|^2|{\cal{M}}|^2\over{E_
{+}E '}}\delta(E'+E_r-1-\nu),
\label{eq:sig}
\end{equation}
where $\phi$(p) and E$_+$ are defined in terms of the
initial quark Dirac momentum wave function
\begin{equation}
\Phi(\vec{p})=\left(\begin{array}{c}
\chi \\
{\sigma\cdot\vec{p}\over{E_{+}}}\chi\end{array}
\right)\phi(p),
\end{equation}
where $\chi$ is a constant Pauli spinor.  In I the
denominator of the small
components of the bound Dirac wave function, E$_{+}$, was
 approximated as a constant parameter of the
  calculation.  In the present application of the
  rest frame model, we use a more realistic bound
  state wave function for which E$_{+}$ is the
    appropriate function of the momentum p.  The Dirac
momentum wave function is normalized so that
\begin{equation}
\int d^3p\Phi^{\dag}\Phi=1,
\label{eq:phi}
\end{equation}
which is different than in I
  where the scalar wave
function $\phi$ was normalized to 1.  For this reason,
Eq.(\ref{eq:sig}) is slightly different than Eq.\ (4) of
I.
With the present normalization, the sum over spins of the
initial bound quark results in the positive energy
projection operator
\begin{equation}
\Lambda={\overline{\not{p}}+\overline{m}\over{ E_{+}}},
\label{eq:lambda}
\end{equation}
which accounts for the 1/E${+}$ in Eq.\ (\ref{eq:sig}).
The
matrix element squared is given by
\begin{equation}
|{\cal{M}}|^{2}=[(4\pi \alpha )^{2}/2Q^{4}][(p'\cdot
k' )(\overline{p} \cdot k) + (p' \cdot
k)(\overline{p}\cdot
k' )-m\overline{m}(k\cdot k' )],
\label{eq:mel}
\end{equation}
where
\begin{equation}
\vec{p}' =\vec{p}
 +\vec{q},\;\hspace{.5in}
p'^{0}=E',\;\hspace{.5in}\overline{\vec{p}}=\vec{p},\;
\hspace{.5in}\overline{p^{0}}=\overline{E},
\label{eq:prime} \end{equation}
$\alpha$=1/137, and $Q^{2}=-q^{2}$.  We neglect
the lepton mass throughout.

      A departure from the free particle matrix
element is the appearance of the quantities
$\overline{E}$ and $\overline{m}$ in Eqs.
(\ref{eq:lambda}-\ref{eq:prime}).
$\overline{E}$ is the energy
component of the four-vector p$^{\mu}$ and, for a
  free quark, would be the quark energy.  However
  for a bound quark, $\overline{E}$ is not the energy,
but   direct calculation with the Dirac equation shows
that $\overline{E}$ can be written as
\begin{equation}
\overline{E}=(E_{+}^{2}+\vec{p}^{2})/2E_+.
\end{equation}
  $\overline{m}$ behaves like an     effective mass of
the bound quark in the initial   state and is given by
\begin{equation}
\overline{m}=(E_{+}^{2}-\vec{p}^{2})/2E_+.
\end{equation}

The angular integration in Eq.\ (\ref{eq:sig}) can be
performed by using the energy delta function to
fix cos$\theta_{pq}$ with the result
\begin{equation}
{d^{2}\sigma\over{d\Omega d\omega'}}=
{\omega'\over{\pi\omega|\vec{q}|}}
\int_{p_{m}}^{p_{M}}{pdp\over{E_{+}}}|\phi|^2
\overline{|{\cal{M}}|^2},
\label{eq:sig2}
\end{equation}
where $\overline{|{\cal{M}}|^{2}}$ is averaged over
$\varphi_{pq}$.  The   limits on the integral, p$_{m}$
and
p$_{M}$, are  the  minimum and maximum momenta,
respectively, for     which the $\delta$ function
argument can vanish.   The minimum momentum $p_{m}$ is
given by the      solution to the quadratic equation that
arises when the energy
conservation equation (1) is solved for p with
$\vec{p}$ and $\vec{q}$
        antiparallel.  In most cases, the maximum
momentum $p_{M}$ is the solution for $\vec{p}$ and
$\vec{q}$  parallel, but there are cases for which
    cos$\theta_{pq}$ never reaches 1 and then p$_{m}$
and p$_{M}$ are two separate solutions  of Eq.\
(1) with cos$\theta_{pq}$= -1.  The
maximum momentum is always greater
than $\nu$ and  can usually be approximated as infinite.

     Equation (\ref{eq:sig2}) gives the cross section for the
scattering of a charged lepton by a bound Dirac quark of
unit charge.  If we assume that the proton is made
up of three Dirac quarks with charges q$_{i}$ such that
$\sum_{i} q_{i}^{2}=1$, that scatter
      incoherently, then Eq.\ (12) can be
      considered the cross section for inelastic
      lepton-proton scattering.  This assumes identical
wave functions for the quarks.  In Ref. [12] we
emphasized that the u and d quarks should
have quite different wave functions in the proton,
corresponding to a diquark-like clustering [19-21].
However, since the two u quarks
      contribute $\frac{8}{9}$ of the cross section and
the d quark only $\frac{1}{9}$, it turns  out  to be a
good approximation, for the proton, to take all quark
wave
functions the same.  In
      lepton-deuteron scattering, where the neutron  is
 included, there is a large dependence on the
d     quark wave function and different d and u quark
wave functions are required to fit scattering  by
deuterons [12,16].

     The assumption of incoherent scattering is
not an essential one, and interference effects
could be calculated if we used a three body wave
function instead of the effective one body wave
function approach described here.  This would
permit comparison with lower Q$^{2}$ data, but
would be a much more complicated calculation.  The
importance of coherence effects depends on the
ratio $|\vec{p}|/|\vec{q}|$ and this is always very small
for the range  Q$^{2}>$4 GeV$^{2}$
considered in this paper.
\vskip 1cm
\no
2.2   THE PROTON STRUCTURE FUNCTIONS F$_{1}$ AND F$_{2}$
\vskip .7cm
\in
     In this paper we fit the lepton-proton
inelastic cross section of Eq.\ (\ref{eq:sig2})
directly to experimental cross sections, but it is still
of
interest to identify the proton structure
functions F$_{1}(x,Q^{2})$ and F$_{2}(x,Q^{2})$,
    where x is the usual Bjorken scaling variable
\begin{equation}
x=Q^{2}/2\nu .
\end{equation}
These can be identified by writing the cross
section as
\begin{equation}
{d^{2}\sigma\over{d\Omega d\omega'}}={\omega'\alpha^2
\over{Q^4\omega\nu}}[4\omega\omega'F_2+Q^2(2\nu F_1-
F_2)].
\label{eq:f}
\end{equation}

In the Bjorken scaling limit, $\nu
\rightarrow \infty$, Q$^{2}\rightarrow \infty$
, with x fixed, F$_{1}$ and F$_{2}$ become functions of
x alone given by
\begin{eqnarray}
F_{1}(x) & = &
2\pi\int_{p_{m(x)}}^{\infty}(p/E_{+})dp|\phi(p)|^2(
\overline{E}+E_r-1+x) \\
F_{2}(x) & = & 2xF_{1}(x),
\label{eq:cg}
\end{eqnarray}
with the minimum momentum depending only on x
through
\begin{equation}
p_{m}(x)=\frac{1}{2}|1-x-M_r^2/(1-x)|.
\end{equation}
  Thus the rest frame model leads to the same scaling
limit as the usual parton model.
There is no log(Q$^2$) scale
breaking in the present model because the final state
quark is taken to be free.  Introducing a QCD final state
interaction between quarks would lead to such log(Q$^2$)
scale breaking that would be important at high Q$^2$.
However, for Q$^2$ less than 20 GeV$^2$, any log(Q$^2$)
effect would be masked by the larger Q$^2$ dependence in
Eqs. (19) and (20) below.

It can be seen from Eq. (15) that quark binding has a
large effect on the structure functions, even in the
scaling limit.  Without binding, E$_+$ would be a
constant (2m), and energy conservation would give
$\overline{E} + E_r = 1$.
Then F$_1$ and F$_2$ would be
simply related to the quark wave function and sum rules
could be derived, similar to those that follow in the
parton model.  However, with binding, these sum rules
cannot be derived in the rest frame model.

     For finite $\nu$ and Q$^{2}$, the approach to
scaling is given to all orders in 1/$\nu$ by
\begin{equation}
F_{i}(x,Q^2)=2\pi(1+2x/\nu)^{-\frac{1}{2}}
\int_{p_m(x,\nu)}^{p_M(x,\nu)}(p/E_{+})dp|\phi(p)|^2f_i
(x,\nu,p). \end{equation}
The integrand functions f$_{i}$ are given by
\begin{eqnarray}
f_1 & = & \overline{E}(1-m/\nu)+\tilde{p}_L[1+(2x-1+E_r-
\overline{E})/\nu]-(\tilde{p}
_L^2/\nu)(2+x/\nu)\nonumber\\
 & & +(p^2/E_{+})(m/\nu)-
p^2x/(\nu^2+2x\nu)+f_2/2\nu,\\
\vspace{.5in}
f_2 & = & 2[\overline{E}(1-E_r)-\tilde{p}_{L}(E_r-
\overline{E}-1)+\tilde{p}_L^2(1-x/\nu)+p^2 x/(\nu+2x)],
\end{eqnarray}
where          $\tilde{p}_{L}$=p$_{L}/\sqrt{1+2x/\nu}$,
and p$_{L}$ is the quark longitudinal momentum given by
\begin{eqnarray}
p_{L} & = & -pcos\theta_{pq} \nonumber\\
      & = & [x-1+E_{r}(1+1/\nu )-(1+M_r^{2}-m^{2})/2\nu
   ]/\sqrt{1+2x/\nu}.
\end{eqnarray}
%footnote{The $\tilde{p}_{L}$ in this paper was called
%$p_L$ in I.
%Also, the F$_1$ defined in Eq. (14) of this paper is
%$\frac{1}{2}$
%the F$_1$ used in I.}
The minimum and maximum momenta for the
integral are now non-scaled functions of x and
$\nu$.  We use the non-scaled structure functions of Eq.
(18) in Eq. (14) to fit the experimental cross sections.
\vskip 1cm
\no
2.3    QUARK WAVE FUNCTION
\vskip .7cm
\in

Because the
integrals in Eq. (18) are over a large range of momentum, the
structure functions do not depend sensitively on the quark wave
function.  It turns out that any relativistic wave function that
falls off in momentum with a power of about $p^5$ can fit the deep
inelastic data, and in I a simple power law momentum wave function
with a constant denominator for the small
components gave a reasonable fit to the SLAC data.  In order to
get a more consistent relativistic relation between the large and
small components, and to more accurately reflect the dominance of
the one gluon exchange potential in the momentum range of our
integrals, we use a Dirac wave function for a Coulomb potential.
While the actual one gluon potential of a single quark in the three
quark system is more like a screened Coulomb, we feel that the use
of a Coulomb potential to connect large and small components is
sufficient in this case.  This is especially true since we do
observe that the structure functions are not sensitive to the exact
details of the wave function.  We find that using bag model wave
functions or relativistic Gaussian wave functions in the integrals
turns out to give a poor fit to the deep inelastic cross sections.

     The  Dirac-Coulomb wave function in momentum
space is given by {\makebox{ Eq. (5)}}, with the scalar part given by
\begin{equation}
\phi^2(p)={b(b+1)(2a)^{2b+1}\Gamma^2(b)sin^2[(b+1)tan^{
-1}
(p/a)]
\over{8\pi^2\Gamma(2b)p^2(p^2+a^2)^{b+1}}}.
\end{equation}
$\phi$(p) is the Fourier transform of the scalar spatial
wave function
\begin{equation}
\psi(r)=Nr^{b-1}e^{-ar}.
\end{equation}
The wave function depends on two parameters: a, the size
parameter, and b which determines the singularity at the
origin of the Dirac wave function.  In momentum space, a
gives the width of the momentum distribution, and b
determines the power of the asymptotic p behavior.  The
denominator of the small components in the momentum wave
function is given by
\begin{equation}
E_{+}  =\frac{p^2b[(1+b)/(1-b)]^{1\over2}}{a-(1+b)p cot[(1+b)
tan^{-1}(p/a)]}.
\end{equation}

     In I it was shown that, if $\phi^2\sim p^{-5}$ at
large p, then F$_2$ would behave like \mbox{(1-x)$^3$} as
x $\rightarrow$ 1, as has been suggested by perturbative
QCD and also noted experimentally.  This corresponds to
b=$\frac{1}{2}$, for which the momentum wave function
could be written in the simpler form
\begin{equation}
\phi^2={3a^2[2a+\sqrt{p^2+a^2}]^2\over{16\pi(p^2+a^2)^3
[a+\sqrt{p^2+a^ 2}]}}
\end{equation}
and
\begin{equation}
E_{+}=\sqrt{3}a+{\sqrt{3}(p^2+a^2)\over{2a+3\sqrt{p^2+a
^2}}}. \end{equation}

     The rest frame model thus depends on two wave
function parameters, a and b, and two masses, the
final state quark mass m and the final state
effective recoil mass M$_r$.  The mass of
the struck quark in the initial state is masked by
the function $\overline{m}$ which acts as an effective
mass in the scattering matrix element.
\vskip 1cm
\newpage
\no
{\bf 3.  The deep inelastic scattering cross sections}
\vskip .7cm
\in

      Most of the accurate data for unpolarized
electron-proton deep inelastic scattering was taken some
time ago at SLAC by several different groups using
different experimental techniques and beams [1-7].
Different methods were also use for
radiative corrections.
 The differences in
radiative corrections and the lack of good relative
normalizations between different experiments led to large
uncertainties in the separation of the two proton
structure
functions and the determination of the ratio
R=$\sigma_L/\sigma_T$.
Recently, these experiments have been reanalyzed by
Whitlow [18] using the same radiative correction
technique for each experiment.  A careful evaluation was
also made of the relative normalizations of the various
experiments.  The Whitlow analysis has resulted in a
consistent set of combined experimental cross
sections measured in six different experiments [1-6]
          over a relatively wide range of kinematical
variables.

Until now, the meeting ground of theory and experiment
for deep inelastic scattering has been at the level of
the structure functions.  That is, the experimentalists
used {\em{ad hoc}} parameterizations of the data to
produce ``experimental" structure functions which were
then used to test theoretical predictions.  But these
structure functions are not truly experimental
quantities. They
     include arbitrary theoretical
constructs in their determination.  This situation has
been much improved with the Whitlow analysis, but the
resulting structure functions still are not direct
experimental quantities.

For this reason, we fit our rest frame model directly to
the Whitlow cross sections so that the theory is tested
directly at the experimental level.  The fit to the
experimental cross sections determines the
parameters of the rest frame model, which can then be
used to determine the separated structure functions
F$_{1}$ and F$_{2}$, the ratio R, and details of the
approach to scaling for Q$^{2}$ in the range 4-20
GeV$^{2}$.  We can also use the model, with its
parameters fixed by the unpolarized fit, to predict the
asymmetry in polarized lepton-proton deep inelastic
scattering and the spin dependent structure functions
g$_{1}$ and g$_{2}$ to compare with  recent
experiments[9,11].

As mentioned earlier, the relative normalization of the
different experiments is of great importance in
separating the two structure functions.  For the small
angle experiments [1-4,6]  at 6-34$^{\circ}$
       there is
enough overlapping deep inelastic data so that
Whitlow could find unambiguous relative normalizations.
However the wide angle experiment [5] at 50$^{\circ}$
and 60$^{\circ}$
         had no deep inelastic points near the smaller
angle experiments.  At the same time, the wide angle
data, precisely because it was in such a different
kinematical region, is the most sensitive for
separating F$_{1}$ and F$_{2}$.  In an attempt to
arrive at some reasonable estimate of the wide angle
normalization relative to the lower angle experiments,
Whitlow  used available elastic scattering comparisons.
But this involved assuming a scaling hypothesis for the
elastic form factors G$_{E}$ and G$_{M}$ that has not
been tested in this region, so that this relative
normalization was more uncertain than the others.

Our procedure in fitting the rest frame model to
experiment has been to treat all normalizations as
experimental quantities to be fit,
 as well as fitting each
experimental point.  That is, we fit to the
combined Whitlow data by minimizing the following
$\chi^{2}$ function
\begin{equation}
\chi^{2}=\sum_{i}^{119}\left({{\sigma_i-\sigma^W_i}
\over{\Delta_i}}\right)^2
+\sum_{j\neq 2}^6\left({{N_{j2}-N^W_{j2}}
\over{\Delta N_{j2}}}\right)^2
+\left({{N_{2}-N^W_{2}}
\over{\Delta N_{2}}}\right)^2.
\label{eq:chi}
\end{equation}
The superscript W refers to
Whitlow, but the cross sections $\sigma^W_i$ in Eq.\
(\ref{eq:chi})
use our
normalization N$_j$ instead of Whitlow's N$^W_j$.  (The
subscript j
refers
to Refs. 1-6 and i refers to each experimental point.)
The
experimental error
$\Delta_{i}$ is that in Whitlow file E.2 with his
normalization error taken out, while $\Delta N_{j2}$
and $\Delta N_2$ are the errors given in
Whitlow table (5.2) as the experimental errors on the
normalization determinations, $N^W_{j2}$ and $N^W_2$,
given in that
table.   Only N$_2$, the normalization for Ref.\ 2
is an absolute    normalization.  Following Whitlow, the
other normalizations are       relative normalizations
N$_{j2}$.   We vary N$_2$ and the
N$_{j2}$ separately.  It is an important feature
of our fit to treat the normalization
errors as normalization errors in this way, and not to
include them in the point to point errors.
\vskip 1cm
\no
{\bf 4.  Results for unpolarized scattering}
\vskip .7cm
4.1  CROSS SECTION FIT
\vskip .5cm
\in

The rest frame model described in section 2  depends on
four parameters: a, the size parameter for the quark wave
function; b, which determines the exponent of the
asymptotic momentum dependence of the wave function;
M$_r$, the effective recoil mass of the spectator quark system;
and m, the final quark mass.  We have fit this model to the
Whitlow combined cross sections by minimizing the
$\chi^{2}$ function in Eq.\ (\ref{eq:chi}) with respect to
the four model parameters a, b, M$_r$, m, and five normalizations.
There is one
absolute normalization and four relative normalizations.
(None of the data points from Ref. 3 survived the cuts
described below.)
In making this fit, we have made cuts on the range of
kinematical variables for which we believe the rest frame
model should apply.  Specifically, we fit only to data in
the ranges \begin{eqnarray}
Q^{2} & > & 4 GeV^{2} \\
W     & > & 3.1 GeV \\
x     & > & 0.3.
\end{eqnarray}
W is the total invariant mass of the final hadronic state
given by
\begin{equation}
W^2=1+[(1-x)/x]Q^2.
\end{equation}

We have determined these ranges by the values of the
respective parameters below which $\Delta
\chi^{2}/\Delta$ DF shows a rapid rise.  The values at
which this happens are reasonable, considering the
expected range of validity of the rest frame model.  The
Q$^{2}$ cut is required so that the incoherence
assumption will be satisfied.  The W cut is required to
leave out direct resonance production (quasi-elastic)
which is not included in the model.  The W cut is not as
necessary for Q$^{2}$ above about 10 GeV$^{2}$,
where the quasi-elastic form factor effectively
eliminates this background.  The x cut is required
because other modes of deep inelastic scattering in
addition to the electron-quark direct scattering
mechanism seem to become
effective at small x.  In I we attributed this mainly to
quark pair production, which is the rest frame
equivalent of the sea in the parton model.

With these cuts, we fit 119 of the points in the
combined Whitlow set, and obtain a $\chi^{2}$ of 198 for
110 degrees of freedom (DF).  Our best fit values are
\begin{eqnarray}
a & = & 393 MeV  \\
b & = & 0.515 \\
M_{r} & = & 0.968 M_p, \\
\label{eq:par}
\end{eqnarray}
with the best quark mass being zero.  The value of
$\chi^2$ increases slowly with increasing quark mass,
reaching 217 at m=300 MeV (with little change in the other parameters), which
still represents a reasonable fit.

     The $\chi^{2}$ for the best fit is
larger than the degrees of freedom.  This is due
to the considerable reduction in the error bars
achieved in the Whitlow combined fit compared to
that of the original experiments.  Earlier fits of
the model to the original data gave a $\chi^{2}$
that was less than the degrees of freedom.  With
the smaller errors, our simple model cannot fit
every nuance of the data, but the overall fit is
quite good.  This is illustrated by two sample cross
section
comparisons in Fig.\ 1.  (There are 34 such
plots that make our cuts, but some have only one or two
points within the cut.)  Figures 1a and 1b show reasonable
agreement, but have relatively high $\chi^2$.
The $\chi^2$/DF, calculated only
for the points within our cuts (square data points), for
Fig.\ 1a is 12/6, and it is 10/7 for Fig. 1b.
    Figure 1a represents the typical way in which the
rest
frame model deviates from experiment beyond our cuts
(circled data points, which were not included in the
fit). The model is usually below experiment for x$<$0.3,
suggesting that other production modes are
effective.  The model is also below the data for
W$<$3.1  GeV where resonance production, which is not
included in the model, is taking place.  In some
cases,  this
disagreement with experiment for W$<$3.1 GeV (at large x)
is not evident in the plot, but, because of the small
error bars, would lead to a large increase in
$\chi^{2}$.

     There are five independent normalizations in the
Whitlow combined fit.  These are the absolute
normalization, N$_2$,   of experiment 2, and four
relative normalizations N$_{j2}$.   Our fit to the
relative normalizations, N$_{j2}$,       j$\neq$5, which
Whitlow could base on overlapping       inelastic data,
all agree with his normalizations.  But  our  fit to the
wide angle relative normalization, N$_{52} $, is
somewhat different than Whitlow's, which was based on
     elastic scattering comparisons and involved some
theoretical assumptions.  Our best fit normalization is
N$_{52}$=1.064,    whereas Whitlow had N$^W_{52}=1.008\pm
.028$. This difference       contributes 4 to our overall
$\chi^2$ in Eq.\ (\ref{eq:chi}).  Whitlow's absolute
normalization  for        experiment 2 also could not be
based on overlapping data.    Our fit to that
normalization is N$_2$=0.919, compared to
N$^W_2=0.981\pm .021$, and contributes 9 to our overall
$\chi^2$.

It is important that the wave function fall
off asymptotically like a
power of p close to $\frac{5}{2}$, which leads to
F$_2(x) \sim (1-x)^3$ as $x \rightarrow 1$.
Gaussian and bag
  model wave functions do not give  reasonable fits to
the deep inelastic cross sections, especially at large x,
 because they do not have this power law momentum dependence.
For our wave function, the large x behavior is given by
F$_2(x) \sim (1-x)^{2(b+1)}$.
     The value b=$\frac{1}{2}$ would correspond in the
Dirac-Coulomb wave
function to a coupling strength  $\lambda$=$\sqrt{3}$/2.
If the two spectator quarks were represented by a point
diquark, this would correspond to a strong coupling
constant
$\alpha_S$=$\frac{3}{4}\lambda$=0.66.  This is a bit
below an estimate
from baryon mass splittings of $\alpha_S$=0.96\cite{jg}.
The smaller $\alpha_S$ found here is a reasonable result
of the spectator quarks being more spread out than a
point diquark.

    The quark wave function is also related to the
initial mass, m$_i$, of the struck quark.  Our value of
a=393 MeV  corresponds to
  m$_i$+S=2a/$\sqrt{3}$=450 MeV,
where S is the
average value of a confining potential, represented in
our model wave function as a constant Lorentz scalar
potential.
This permits a reasonable range of quark masses and
confining potentials.

 More complicated  wave functions
could be used with more realistic spectator quark
distributions and a linear confining potential.  We do
not feel that this would appreciably improve the fit, although it
could modify the quark model size and mass parameters.
\vskip .7cm
\no
4.2  STRUCTURE FUNCTIONS
\vskip .5cm
\in

  Once the parameters of the model have been
set by fitting to the experimental cross-sections,
we can calculate the structure functions F$_{1}$
  and F$_{2}$ as functions of the invariants.
Figure 2 shows F$_2$
as a function of Q$^2$ for several values
of  x, along with
 the Whitlow determinations (data points)
of F$_{2}$.  We
     emphasize that this is not a comparison of theory
with experiment, but a comparison of two different
methods of extracting the structure functions
from the same experimental cross sections.  The points plotted
exclude the region with
W less than 3.1 GeV  for which there would be quasi-elastic
resonance production, which is not included in our model.
The agreement in x and Q$^2$ is reasonable,
except for the wiggle in the Whitlow extraction at x=0.35,
which would be hard to reproduce in any simple model.

     Another quantity of interest is the ratio
R=$\sigma_{L}/\sigma_{T}$, which is related to F$_{1}$
and F$_{2}$ by
\begin{equation}
R=(1+2x/\nu)F_{2}/2xF_{1}-1.
\end{equation}
R should approach zero in the scaling limit.  The
early experimental determinations [2,4,6] had
large errors because the experiments covered a
small range (6-30$^{\circ}$)
        of angles and the
different experiments had uncertain relative
normalizations.  The Whitlow analysis
of the early experiments readjusted the relative
normalizations to make them compatible where they
overlap, and have added experimental cross
sections [5] at 50$^{\circ}$ and 60$^{\circ}$,
        but with a more
uncertain relative normalization.

    We compare our
determination of R with that of Whitlow (data
points) in Fig.\ 3.
These, also, represent two different extractions of R
from the same
cross sections.  There is reasonable agreement, although
there is
considerable scatter in the Whitlow R extraction.  The
agreement
continues even into the low W region.
\vskip 1cm
\no
{\bf 5.  Polarized deep inelastic muon scattering}
\vskip .7cm
5.1 POLARIZATION EXPERIMENTS
\vskip .5cm
\in

     The polarized deep inelastic scattering experiments [8-11] have generally
measured the  asymmetry in the
scattering of charged longitudinally
polarized leptons by longitudinally polarized protons.
(Ref.\cite{slac2} also measured transverse asymmetry.)
Specifically, they have measured the quantity
\begin{equation}
A={{{d^2\sigma^{\uparrow\downarrow}\over{d\Omega
d\omega'}}- {d^2\sigma^{\uparrow\uparrow}\over{d\Omega
d\omega'}}}\over{{d^2\sigma^{\uparrow\downarrow}
\over{d\Omega d\omega'}}+{d^2\sigma^{\uparrow\uparrow}
\over{d\Omega d\omega'}}}}
\label{eq:asym}
\end{equation}
where the arrows refer to the initial lepton and proton
spin projections along the incident lepton direction in the
laboratory.  The denominator of Eq.\ (\ref{eq:asym}) is
just twice the unpolarized cross section given by
Eq.\ (\ref{eq:sig2}).

    Dimensionless  spin dependent structure functions
g$_1$ and g$_2$ can be defined in terms of the
numerator of Eq.\ (\ref{eq:asym}) by
\begin{eqnarray}
{\cal{N}} & = & {d^2\sigma^{\uparrow\downarrow}
\over{d\Omega d\omega'}}-
{d^2\sigma^{\uparrow\uparrow}
\over{ d\Omega d\omega'}}\nonumber\\
& = &
{4\alpha^2\omega'\over{Q^2\nu}}\left([2-
{\nu x\over{\omega^2}}]g_1(x,Q^2)-
{\nu\over{\omega}}[g_1(x,Q^2)+
{2x\over{\nu}}g_2(x,Q^2)]\right).
\label{eq:polg}
\end{eqnarray}

     The measured asymmetries can be given in terms of
the virtual photon asymmetry
\begin{equation}
A_1={{\sigma_{1\over2}-
\sigma_{3\over2}}\over{\sigma_{1\over2}+
\sigma_{3\over2}}},
\end{equation}
where $\sigma_J$ is the virtual photon-proton
absorption cross section with total spin projection J
along the photon momentum.
A$_1$ is related to the structure functions by
\begin{equation}
A_1=(g_1-2xg_2/\nu)/F_1.
\end{equation}
\vskip .7cm
\no
5.2   REST FRAME MODEL FOR POLARIZED SCATTERING
\vskip .5cm
\in

     The rest frame model treats lepton-quark
scattering which gives ${\cal{N}}_{quark}$, and this
must be related to ${\cal{N}}_{proton}$.  There is a
depolarization of the quarks with respect to the
proton polarization.  This depolarization follows from
the proton spin wave function
\begin{equation}
P\uparrow>={1\over{\sqrt{6}}}
(2\uparrow\uparrow\downarrow-
\uparrow\downarrow\uparrow-
\downarrow\uparrow\uparrow),
\end{equation}
where the order (uud) is understood for the valence
quarks constituting the proton.  The polarized lepton-
proton cross sections are the sum of the lepton-quark
cross sections, weighted by the squares of the quark
charges.  This leads to

\begin{eqnarray}
\sigma^{\uparrow\uparrow}_{proton} & = &
\frac{1}{9}(7\sigma^{\uparrow\uparrow}
+2\sigma^{\uparrow\downarrow})_{quark} \\
\sigma^{\uparrow\downarrow}_{proton} & =&
{1\over{9}}(2\sigma^{\uparrow\uparrow}
+7\sigma^{\uparrow\downarrow})_{quark},
\end{eqnarray}
and, from Eq.\ (\ref{eq:asym})
\begin{equation}
A_{proton}=\frac{5}{9}A_{quark}.
\end{equation}
As in the unpolarized case, we have assumed the same
momentum wave function for the u and d quarks.

     The polarized lepton-quark cross section
calculation proceeds like the unpolarized calculation,
but with spin projection operators inserted into the
initial lepton and quark matrix elements. The spin
projection operators are given by
\begin{equation}
\Lambda_{\uparrow}={1\over
2}( 1+\gamma^5\gamma_{\mu}W^{\mu}),
\end{equation}
where W$^{\mu}$ is the covariant spin vector for
polarization in the $\hat{k}$ direction, given by
\begin{equation}
\overline{m}W^{\mu}_p=[\vec{p}\cdot\hat{k},
(\vec{p}\cdot\hat{k})(\vec{p}/E_{+})+\overline{m}\hat{k}]
\end{equation}
for the initial struck quark, and
\begin{equation}
m_LW^{\mu}_{L}=(\omega,\vec{k})
\label{eq:Lspin}
\end{equation}
for the lepton. The lepton mass m$_L$ has been
neglected on the right hand side of
Eq.~(\ref{eq:Lspin}).

     The resulting matrix element squared for polarized
lepton-quark scattering is
\begin{eqnarray}
|{\cal{M}}^{\uparrow\uparrow}|^2 & = & [(4\pi\alpha)^2/2Q^4]
\verb+{+[(k'\cdot p')(k\cdot\overline{p})
+(k'\cdot\overline{p})(k\cdot p')
-m\overline{m}(k\cdot k')]\nonumber\\
 & & +m_L\overline{m}[(W_L\cdot
W_p) (-q\cdot p')+(W_L\cdot p')(W_p\cdot q)]\nonumber\\
 & & +m_Lm[(W_L\cdot W_p)(-q\cdot\overline{p})+(W_L\cdot
\overline{p})(W_p\cdot q)]\verb+}+.
\label{eq:pmel}
\end{eqnarray}
$|{\cal{M}}^{\uparrow\downarrow}|^2$ is given by
Eq.\ (\ref{eq:pmel}) with W$_p$ replaced by -W$_p$.
Then, repeating the steps of the unpolarized
calculation leads to polarized cross sections from
which g$_1$ and g$_2$ can be identified using
Eq.\ (\ref{eq:polg}).
\begin{equation}
g_i(x,Q^2)=(\frac{5}{9})2\pi(1+2x/\nu)^{-
\frac{1}{2}}\int_{p_m}^{p_M}(p/E_{+}) dp\phi^2h_i,
\label{eq:g}
\end{equation}
with the factor $\frac{5}{9}$ coming from the depolarization in
Eq. (44).
The integrand functions are given by
\begin{eqnarray}
h_1 & = & \overline{m}+\tilde{p}_L(1-
\overline{m}/\nu)+(\tilde{p}_L^2/E_{+})[1-
(E_r+E_{+}-1)/\nu]\nonumber\\
 & & +(\tilde{p}_T^2/2E_{+})
[1+(E_{+}+E_r-1+2x)/\nu]\nonumber\\
 & & +(m/\nu)(\tilde{p}_L+\tilde{p}_L^2/E_{+}-
\tilde{p}_T^2/2E_{+}) \\
\vspace{.3in}
h_2 & = &
{1\over{2x}}[\overline{m}(E_r-1+2x)+\overline{m}
\tilde{p}_L(1+2x/\nu)]\nonumber\\
 & & +(p_T^2/2E+)(E_{+}+E_r-1+2x)\nonumber\\
 & & - {m\over{2x}}[\overline{E}+\tilde{p}_L(1+2x/\nu) -
p_T^2/2E_{+}]-h_1.
\end{eqnarray}
The parameters of the model have been set by the fit to
the unpolarized cross sections, so that the
polarization predictions have no adjustable parameters.
It can be seen again, from Eqs. (50) and (51), that, due to binding
effects, the connection of g$_{1p}$ to the quark wave
function is not simple enough to derive parton model type
sum rules for the integral of g$_{1p}$, even in the
scaling limit.

\vskip .7cm
\no
5.3    POLARIZATION RESULTS
\vskip .5cm
\in

     The virtual photon asymmetry A$_1$ is plotted in
Fig.\ (4).
The experimental points are from the SMC muon experiment\cite{smc}
(solid points) and the SLAC E143 electron experiment\cite{slac2}
 (open points).   The square points in each case are within the
x$>$0.3, W$>$3.1 GeV, Q$^2>$4 GeV$^2$ range of the rest frame model.
The SMC data was taken at much higher Q$^2$ than the E143 data.
The  curve is the rest frame prediction for $A_1$, calculated
at each E143 point for the actual Q$^2$ at that point.
  The rest frame model is in reasonable agreement (We discuss the jump in the
E143 asymmetry around x=0.4 later.)
with these experiments in the range x$>$0.3, but,
for x$<$0.3 the predicted asymmetry is much above
that of  both experiments.  This is because the
calculated unpolarized cross section in the denominator
of the asymmetry definition Eq.\ (37) is too
small in the rest frame model for x$<$0.3.

     In Fig.\ 5a
our predictions for g$_{1p}$,
which is related only to the difference of cross
sections, is shown along with the SMC
experimental values.  Our calculations for g$_{1p}$
are made at the experimental value of Q$^2$ for each data
point.  The solid triangle on the x axis shows the x value below which
Q$^2$ is less than 4 GeV$^2$.

   The rest frame prediction of this paper (dashed curve),
gives a $\chi^2$ of 24 for the 12 SMC data points, including
all the points on the curve.
  In Ref. \cite{mi}, it was shown that different d and u
quark wave functions are required to simultaneously
fit unpolarized deuteron and proton cross sections.
Using those different d and u wave functions has little
effect on the unpolarized proton cross section, but decreases
the proton g$_1$ at small x (solid curve), improving
$\chi^2$ to 14 for the 12 data points in Fig. 5a.
  We interpret this good agreement, even below
x=0.3, to mean that the higher diagrams left out of our
calculation do not contribute much to the polarization
and their effect cancels out of the difference between
polarized cross sections.  The improvement of the prediction for
different u and d quark distributions shows that the polarized
scattering is a more sensitive test of SU(6) breaking in the proton than is the
unpolarized scattering.

Figure 5b shows the SLAC E143 measurement of g$_{1p}$, along with the
rest frame predictions for the same (dashed curve), and different
(smooth curve) u and d quark distributions.  The E143 g$_{1p}$ was evaluated at
a constant Q$^2$=3.0 GeV$^2$ for each x, and we use that Q$^2$ in our
calculation for this figure.   Again, different u and d distributions
give a better prediction for g$_{1p}$.

 In Fig. 5b, the solid triangle on the x axis indicates the point below which
Q$^2<$4 GeV$^2$, and the open triangle the point above which W$>$3.1 GeV$^2$.
We attribute the jump in the E143 g$_{1p}$ around x=0.4 to quasi-elastic
production of resonances at these lower values of W.
The lowest x point in the jump at x=0.416 has W=3.3, just above our chosen
cutoff of W=3.1 GeV.  The five higher x points all have W$<$3.1 GeV.
In fitting the unpolarized cross sections, the W cutoff was chosen somewhat
arbitrarily, since resonance production does not set in at one particular
energy.  There is only one SMC point in this x range.  We show that by the open
circle at x=0.48.  This SMC point has W=8 GeV and Q$^2$=58 GeV$^2$.  By
contrast, the E143 point at x=0.47
has W=3.04 GeV and Q$^2$=7.4 GeV$^2$.  It  will be interesting to see if future
data in this x region comes down to the SMC level as W and Q$^2$
are increased.

Figure 6 shows the rest frame prediction for g$_{2p}$.   It is seen to be very
small until x gets below about 0.01.
  Even though g$_2$
gets quite large at small x, it is multiplied by x/$\nu$ in
Eq.\ (\ref{eq:polg}) for $\cal{N}$, and does not affect
the extraction of g$_1$ from longitudinal asymmetries.

\no
\vskip 1cm
{\bf 6.  Discussion}
\vskip .7cm
\in

     The version of the rest frame model presented here
is quite simple in that we have used a simple two
parameter wave function, and a delta function
approximation of the spectator quark relative momentum
distribution.  We have also calculated only the lowest
order diagram, which restricts our model to the
range of x$>$0.3.  In that range, the cross
section fit looks quite good, as indicated by Fig.\ 1.

The model could be extended in a number of
ways:
\begin{enumerate}
\item A more sophisticated wave function could be
used.  The fit is already good enough that we do not feel
that this is warranted, and a better wave function would
probably not extend the range of validity, although it
could affect the best fit quark model parameters.

\item The additional process of quark pair production by
the virtual photon, followed by quark (or
 anti-quark) scattering via QCD could be included.  This
could extend the validity of the model below x=0.3.  The
quark pair diagram in the rest frame
corresponds to the quark-antiquark sea in the infinite
momentum frame.  That is, the rest frame model does not
have a $q-\overline{q}$ sea, but the pair production
diagram would contribute.  While at infinite momentum,
the quark pair diagram (a so called z-graph) does not
contribute, but a $q-\overline{q}$ sea is generally
introduced at low x.  In each case, the simple valence
quark model does not contribute enough at low x.

\item Gluon bremsstrahlung and final state interaction
between the quarks via one gluon exchange could be
included.  This would
lead to the logarithmic Q$^2$ dependence,
characteristic of QCD.  Without this
final state interaction, the simple model used here has
pure scaling in x as Q$^2$ becomes large.

\item Different u and d quark wave functions should be
used. This is not too important for the proton where the
d quarks contribute only ${1\over{9}}$ to the cross
section, but is needed to describe the deuteron and the
x dependence of the ratio
F$_{2n}$/F$_{2p}$ [12,16].  We have also shown in Fig. 5 that different u and d
quark distributions improve the prediction for g$_{1p}$.
\end{enumerate}
\vspace{.15in}

     The rest frame model used here is very much like the
parton model in its application to experiment.  Our
structure functions have
the same experimental application as those of the parton
model. However, the proton looks quite different in the two models.
In the rest frame model, the proton wave function is
simple, with only the three valence quarks.  In the
parton model, implemented in the infinite momentum
frame (or on the light cone), the dynamics is simpler,
but the wave function more complicated in having an
explicit gluon component and a $q-\overline{q}$ sea.
Since the results of each model can be presented in terms
of Lorentz invariants, they must be equivalent in some
sense.   We believe that their equivalence lies in the
quark pair diagrams in the rest frame model
corresponding to the sea at infinite momentum, and the
gluons at infinite momentum corresponding to the Lorentz
transformation of the internal energy of the rest frame
quarks.

In its polarization predictions, the
rest frame model seems particularly clearer.  Our success
in predicting the measured g$_{1p}$, even at low x,
indicates that the complicating diagrams in the rest
frame do not contribute to the polarization, which is
accounted for completely by the valence quarks.  While in
the parton model, the sea and  the gluon component both
seem to be highly polarized so that there is no simple
understanding of the origin of the proton spin.  One
could say that the rest frame model is lucky in that the
mechanisms left out of the model (at this stage) do not
seem to contribute to the polarization, while the parton
model is unlucky in this respect.

\vskip .5cm
\no
{\bf 6.  Conclusion}
\vskip .3cm
\in

     The simple version of the rest frame model
presented here has given a reasonable fit to the
unpolarized deep inelastic cross sections.  With the
model based on this unpolarized fit, we find good
agreement with polarization
measurements of the spin dependent stucture function
g$_{1p}$ of the proton.
  We conclude from this that deep
inelastic scattering can be understood by looking at the
proton, in its rest system, as composed of three
relativistically bound valence quarks, with the spin of
the proton carried completely by the valence quarks.
%We also note that parton model sum rules cannot be derived
%in the rest frame because of binding effects.
\vskip 1cm

One of us (JF) would like to thank Temple University for
a Reseach Leave and the Lady Davis Fellowship Trust for a
Lady Davis Fellowship at the Technion where part of this work
was performed.

\

\no
\vskip .5cm
{\bf Figure Captions:}
\vskip .3cm

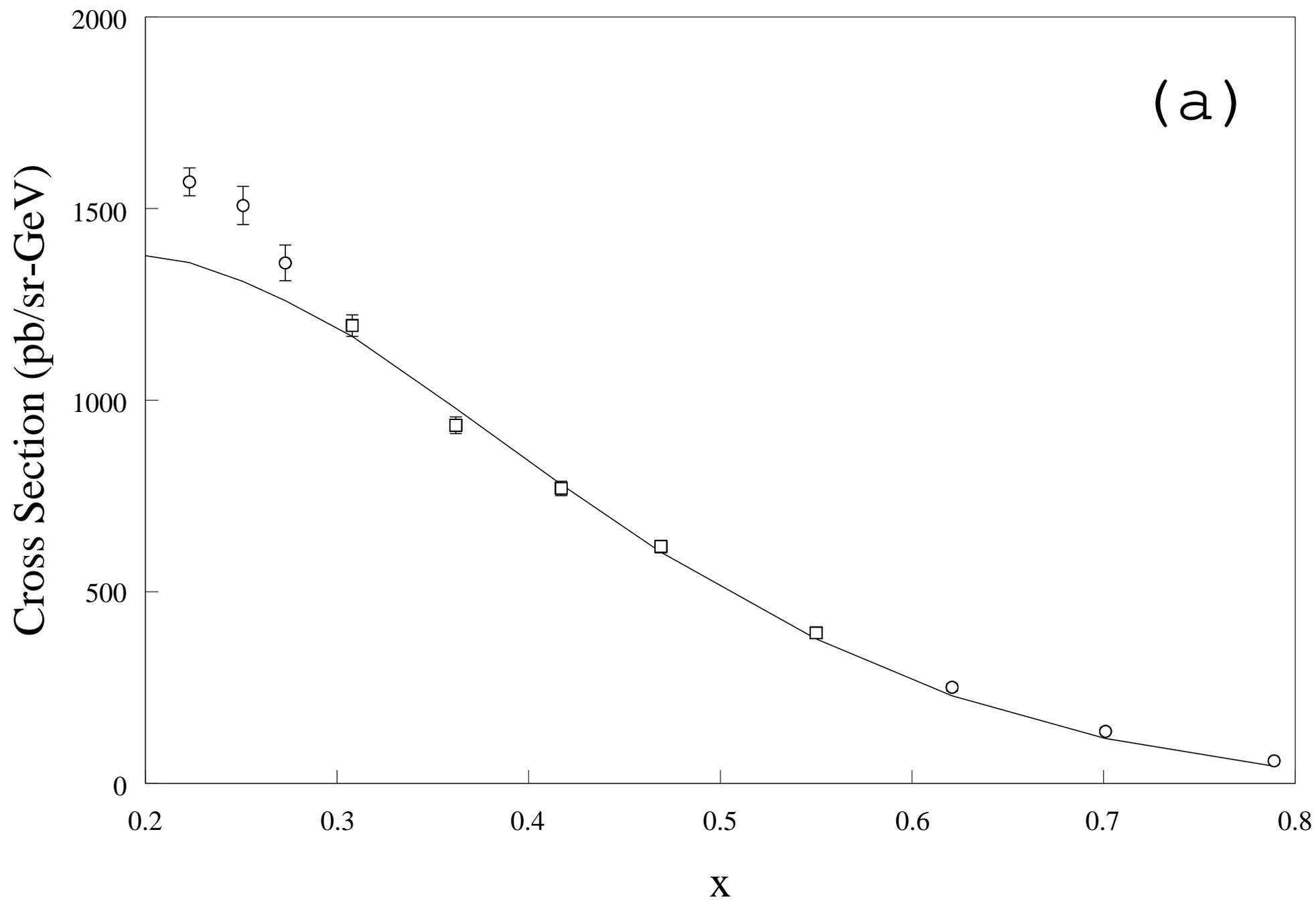
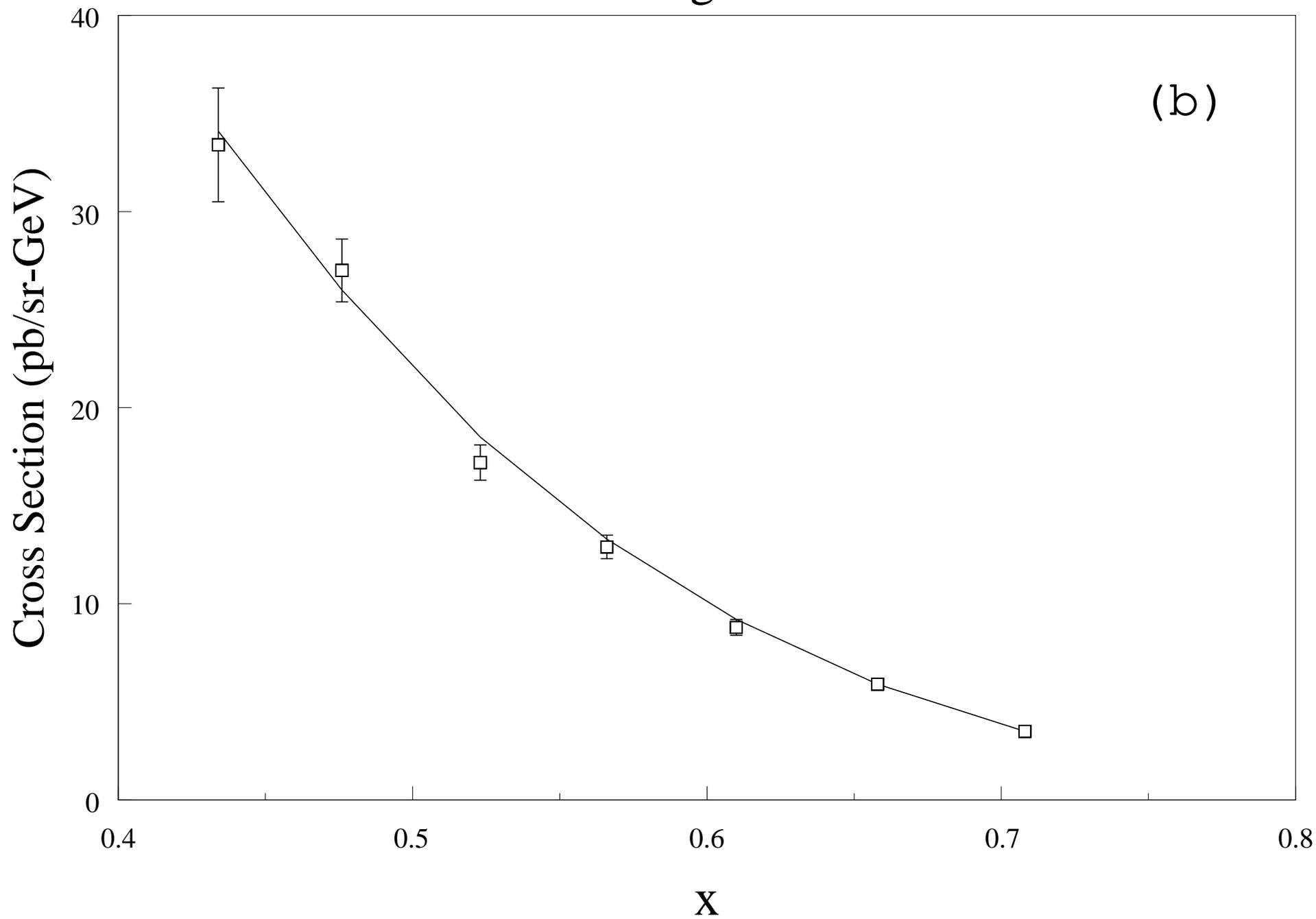
FIG.\ 1. Deep inelastic electron proton cross sections.
The data points are from Whitlow, Ref.\ 18.  The circle
points are beyond our cuts and were not included in the
$\chi^2$ fit.  (a)  Scattering angle $\theta=18^{\circ}$
with incident electron energy $\omega=17.0$ GeV.
(b) $\theta=50^{\circ}$,
     $\omega=19.5$ GeV.
\vskip .3cm
FIG.\  2.  The structure function F$_2$(x,Q$^2$).  The
data points are Whitlow's extraction of F$_2$ in Ref.\ 18.
\vskip .3cm
FIG.\ 3.  R=$\sigma_L/\sigma_T.$  The data points are from
Whitlow's extraction of R in Ref.\ 18.  (a) x=0.40.  (b)
x=0.55.  (c) x=0.70.
\vskip .3cm
FIG.\ 4.  The virtual photon asymmetry A$_1$.   The solid
data points are from SMC, Ref.\ 11, and the
open points are from E143, Ref.\ 9.  The solid curve is the
rest frame prediction for different u and d quark distributions, and the
dashed curve is for equal u and d quark distributions.
\vskip .3cm
FIG.\ 5.  The spin dependent structure function g$_1p$.
The experimental points in (a) are from SMC, Ref.\ 11,
and in (b) from E143, Ref.\ 9.
The dashed curve is for similar, and the solid curve for different
u and d quark distributions.
\vskip .3cm
FIG.\ 6.  The rest frame prediction for the spin dependent structure function
g$_{2p}$.
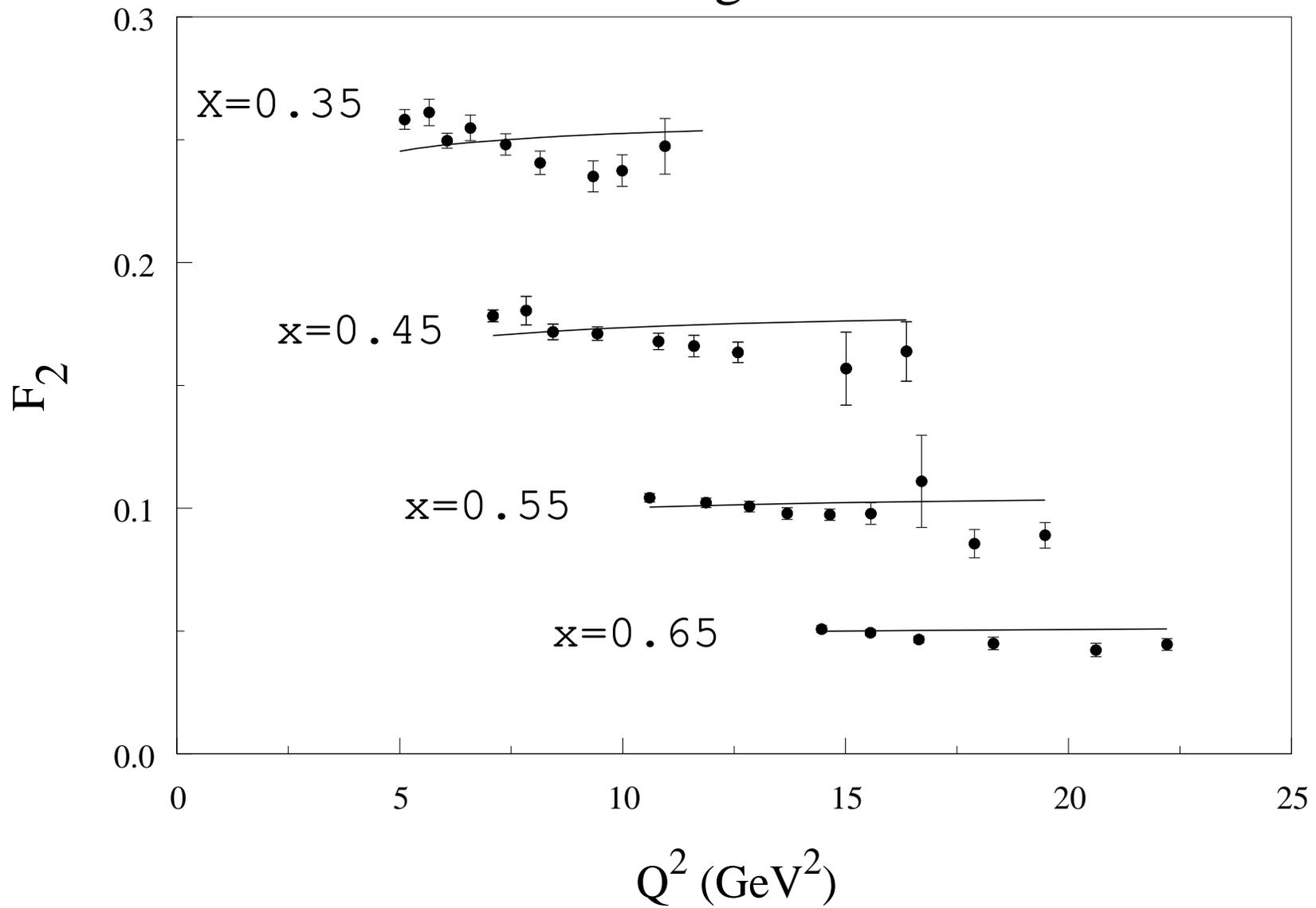
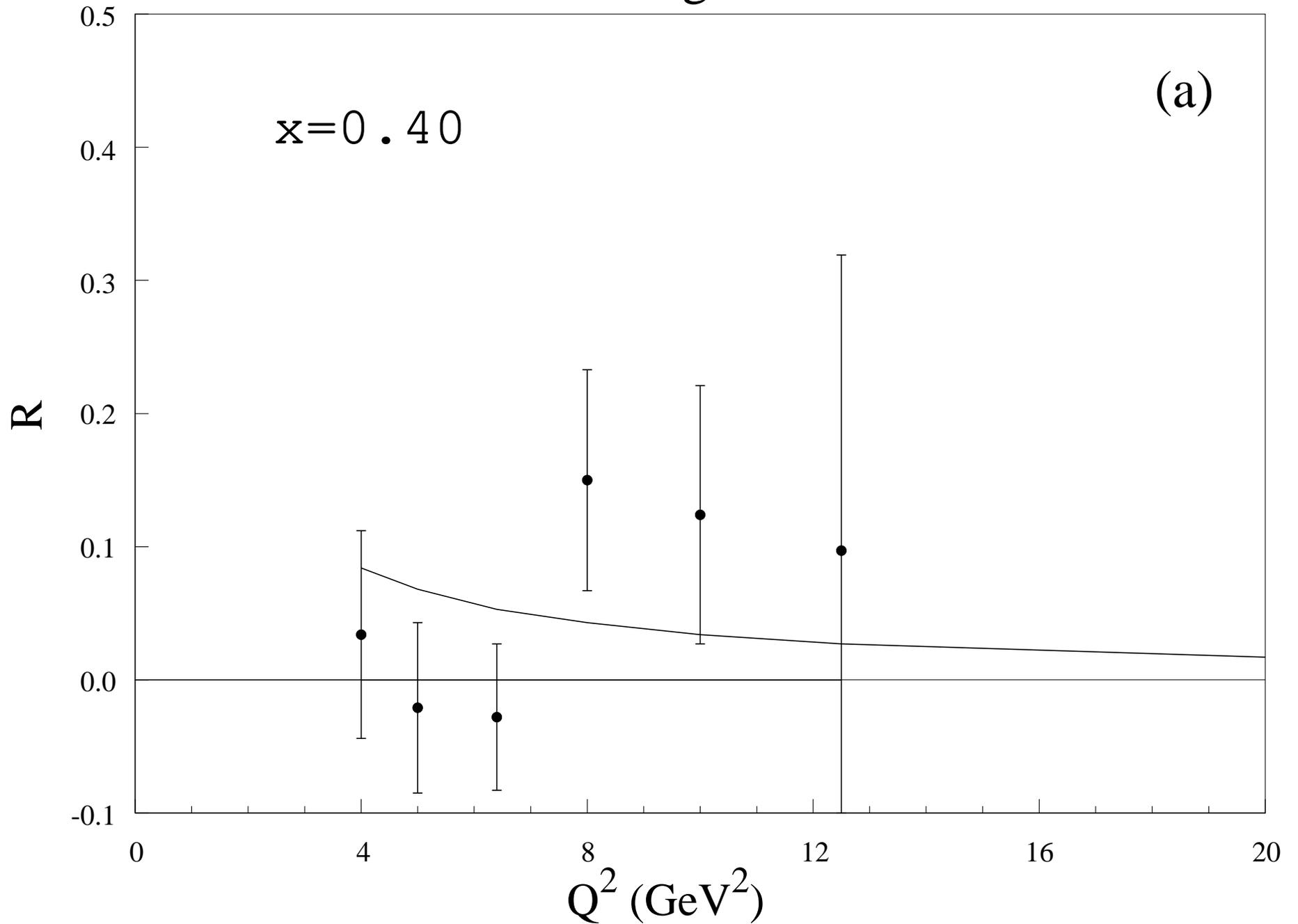
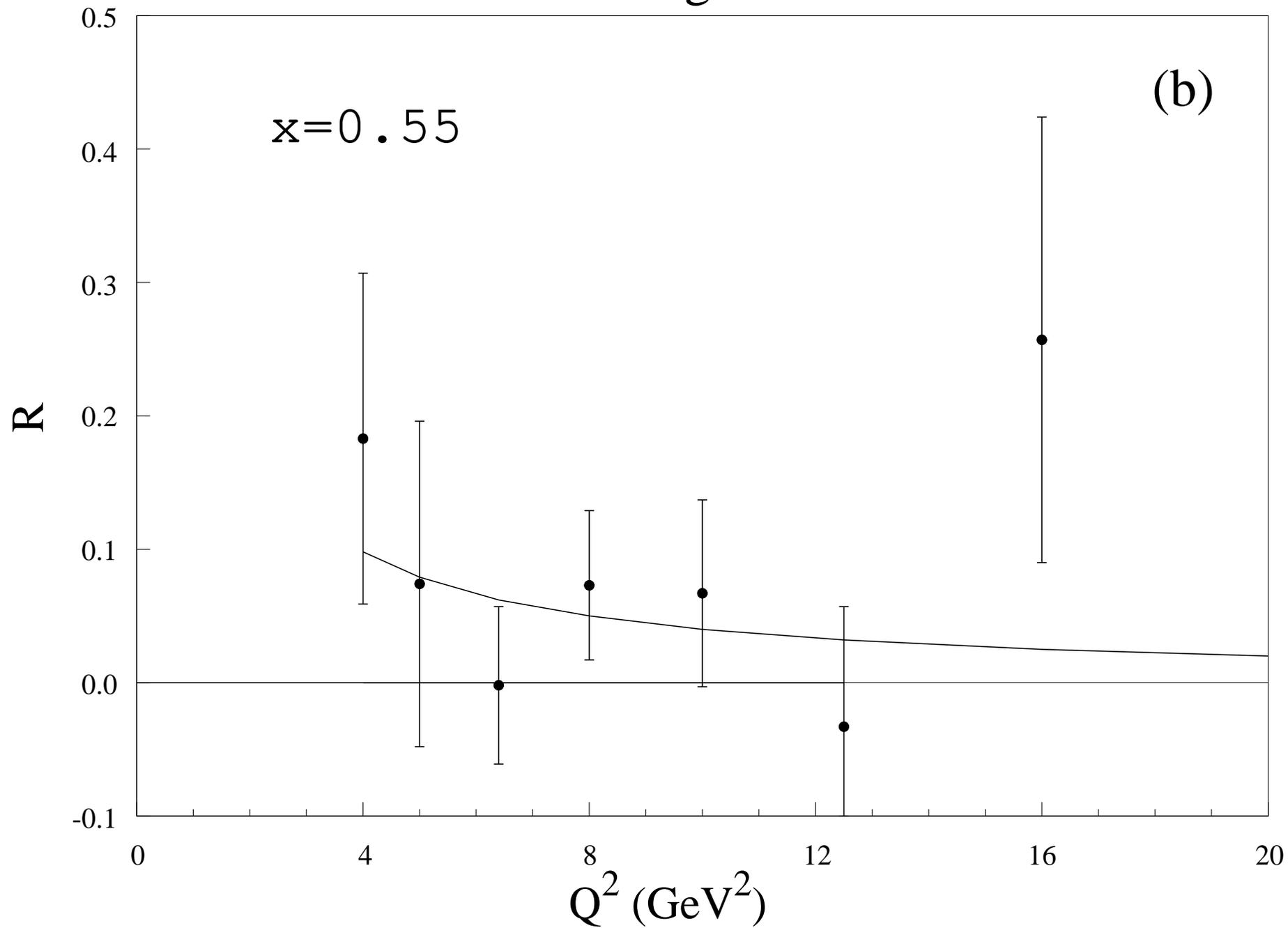
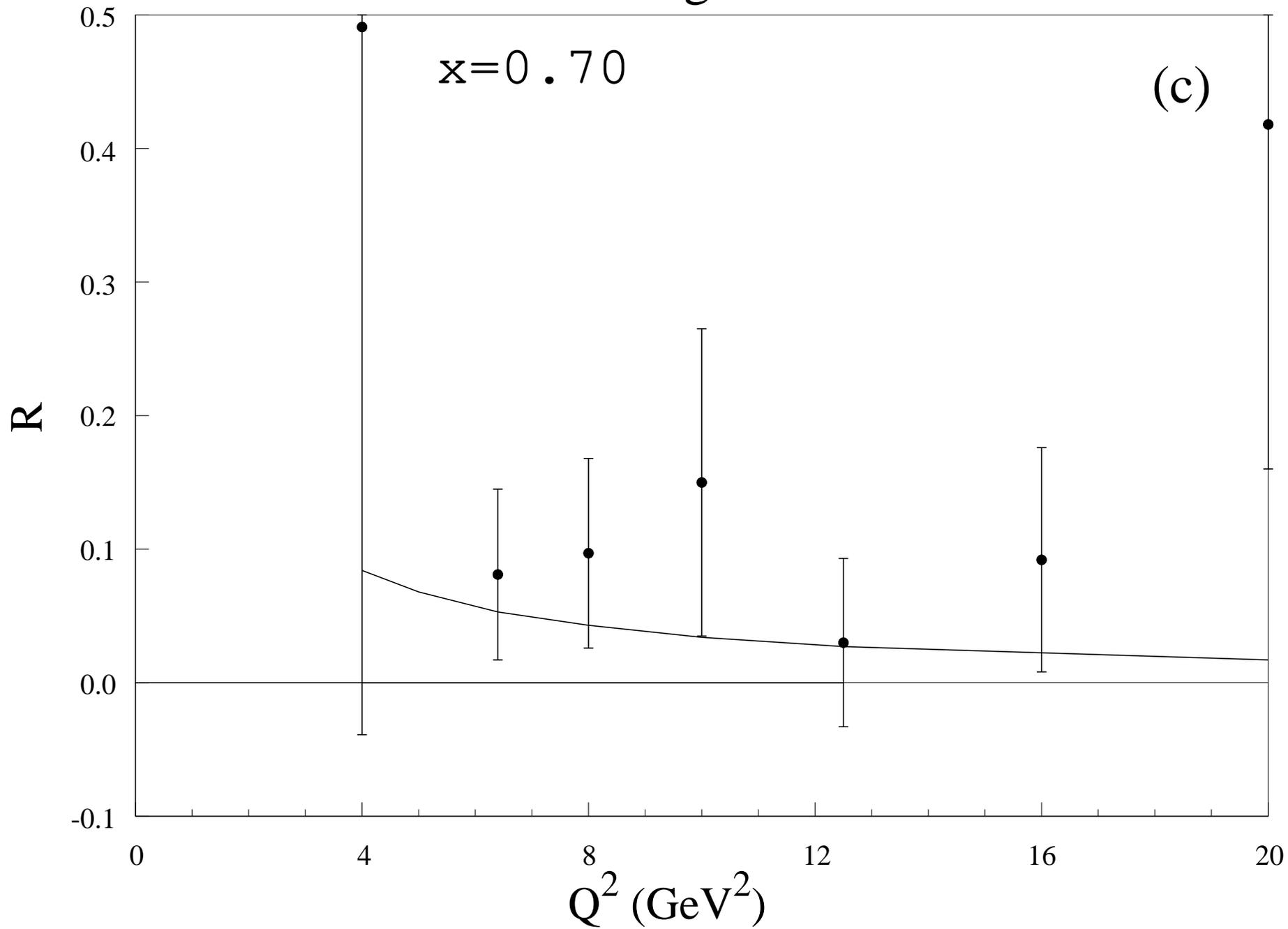
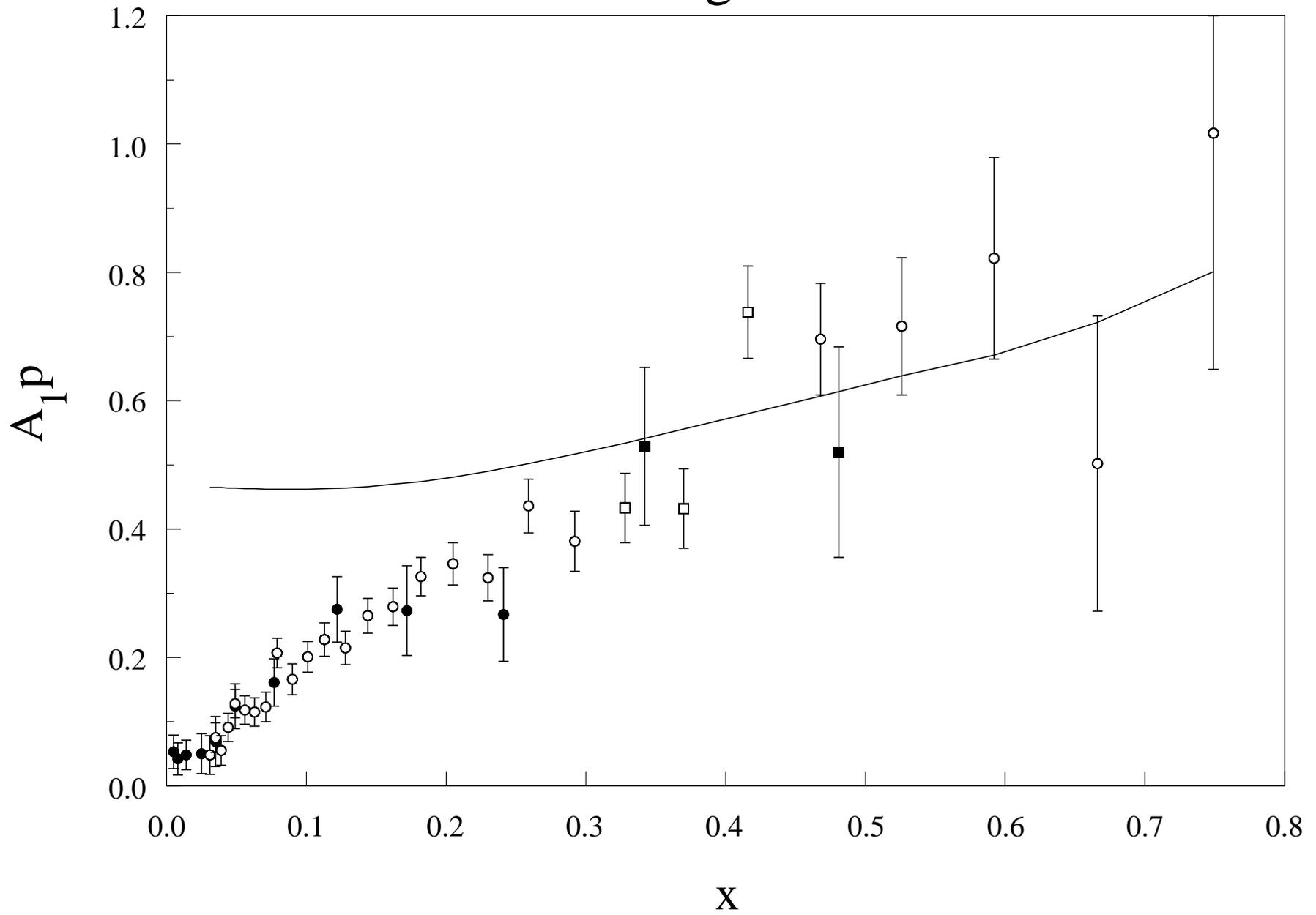
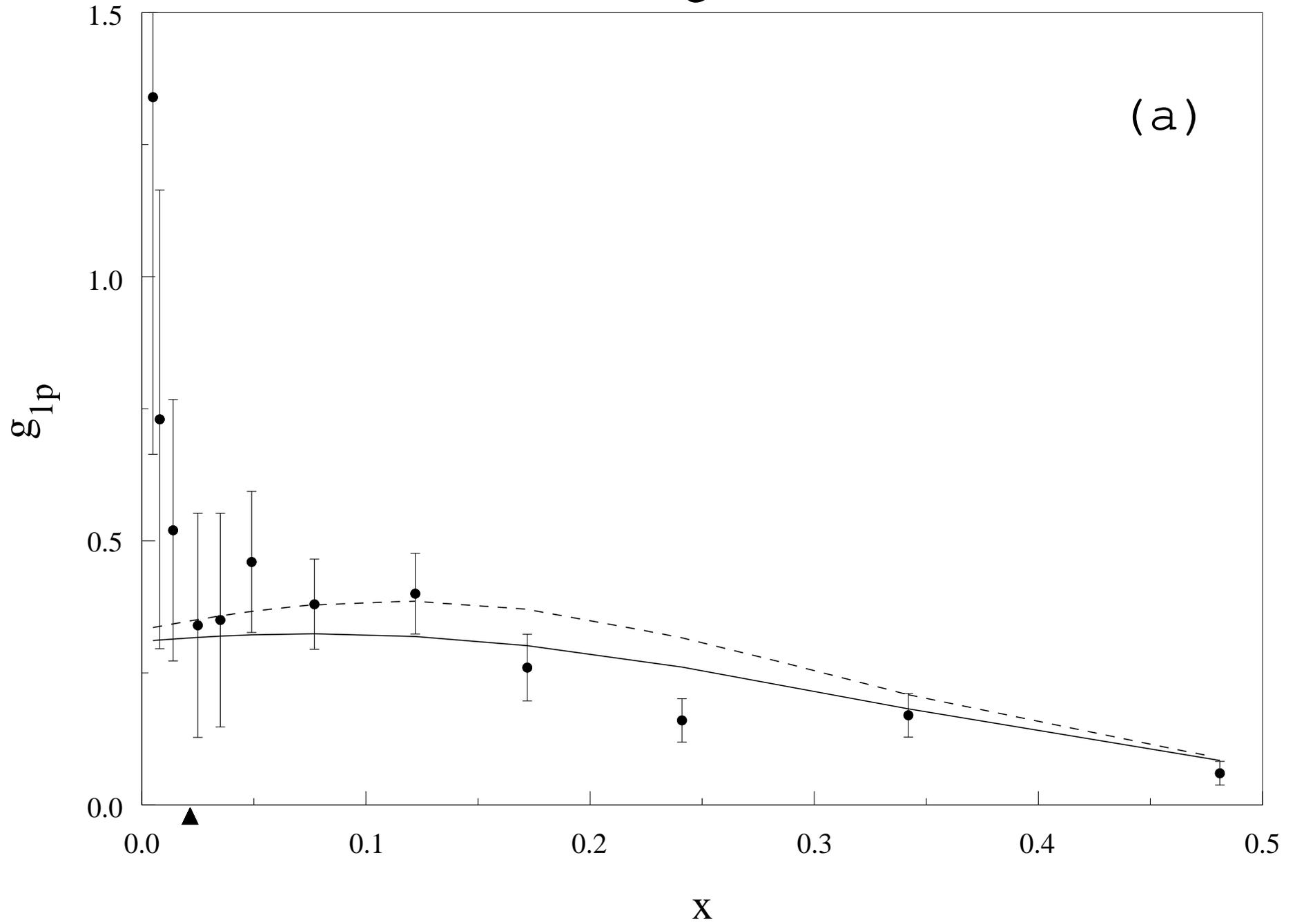
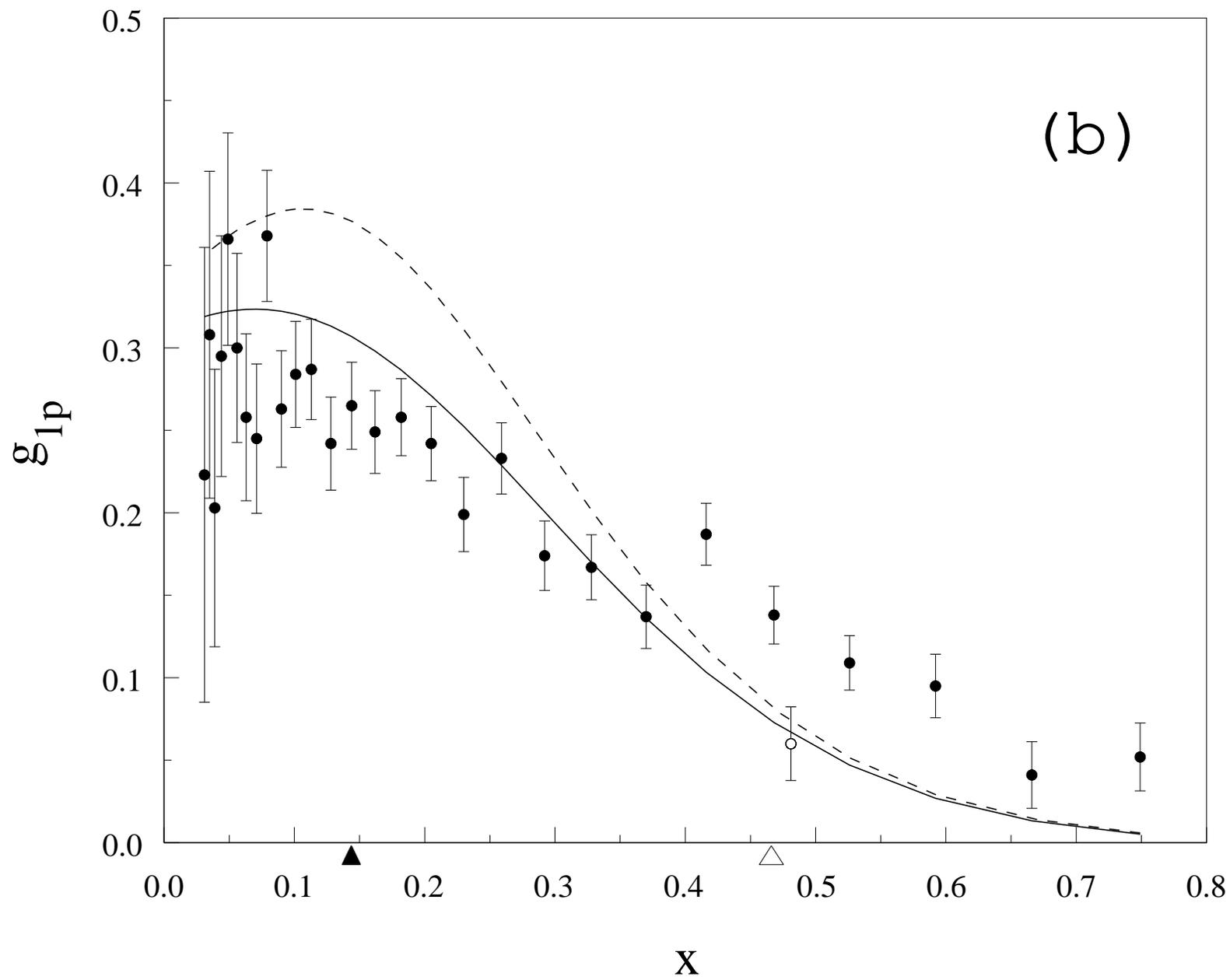
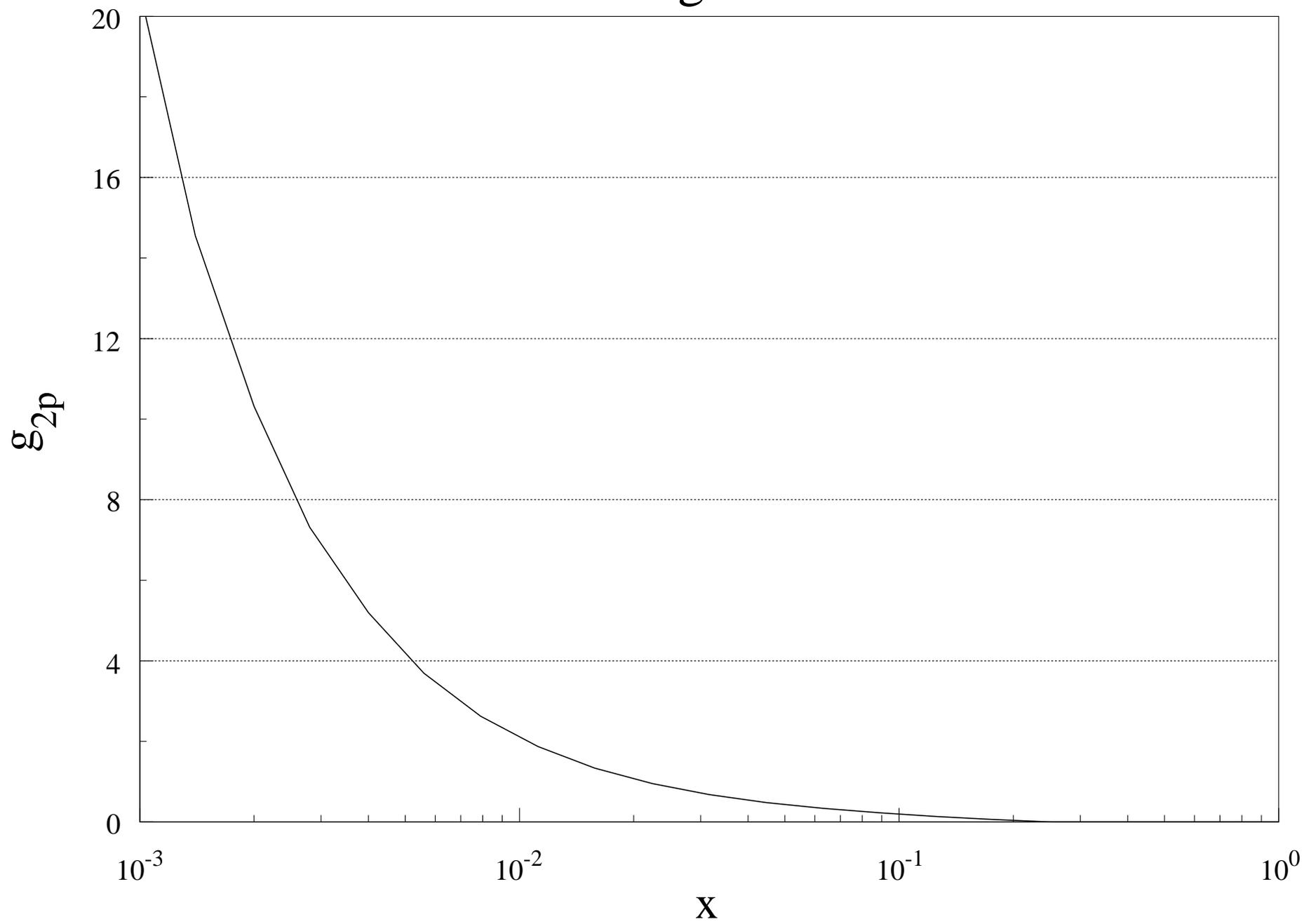
\end{document}